\documentclass[5p,authoryear]{elsarticle}

\usepackage{natbib}

\usepackage{epsfig}

\usepackage{amssymb}
\begin{document}

\begin{frontmatter}

\title{Evolution of Magnetic Fields and Cosmic Ray Acceleration in Supernova Remnants}
\author[address1,cor]{K.M. Schure}
\address[address1]{Astronomical Institute, University of Utrecht,
             P.O. Box 80000, NL-3508 TA Utrecht, The Netherlands}
\ead{K.M.Schure@phys.uu.nl}

\author[address1]{J. Vink}
\author[address1]{, A. Achterberg}
\author[address1,address2,address3]{, R. Keppens}

\address[address2]{Centre for Plasma Astrophysics, K.U. Leuven, Celestijnenlaan 200B, 301 Heverlee, Belgium}
\address[address3]{ FOM-Institute for Plasma Physics ``Rijnhuizen'', P.O. Box 1207, NL-3430 BE Nieuwegein,The Netherlands}

\begin{abstract}
Observations show that the magnetic field in young supernova remnants (SNRs) is significantly stronger than can be expected from the compression of the circumstellar medium (CSM) by a factor of four expected for strong blast waves. Additionally, the polarization is mainly radial, which is also contrary to expectation from compression of the CSM magnetic field. Cosmic rays (CRs) may help to explain these two observed features. They can increase the compression ratio to factors well over those of regular strong shocks by adding a relativistic plasma component to the pressure, and by draining the shock of energy when CRs escape from the region. The higher compression ratio will also allow for the contact discontinuity, which is subject to the Rayleigh-Taylor (R-T) instability, to reach much further out to the forward shock. This could create a preferred radial polarization of the magnetic field. With an adaptive mesh refinement MHD code (AMRVAC), we simulate the evolution of SNRs with three different configurations of the initial CSM magnetic field, and look at two different equations of state in order to look at the possible influence of a CR plasma component. The spectrum of CRs can be simulated using test particles, of which we also show some preliminary results that agree well with available analytical solutions.

\end{abstract}


\end{frontmatter}

\parindent=0.5 cm

\section{Introduction}

Supernova remnants (SNR) are sites where particles are believed to be accelerated up to energies of $10^{15}$~eV. Cosmic rays (CRs) scatter off local magnetic field perturbations and are accelerated by the first order Fermi mechanism, also referred to as diffusive shock acceleration (DSA). In order to accelerate CRs to sufficiently high energies and explain the non-thermal X-ray synchrotron emission that is confined to a small region around the blast wave, the magnetic field needs to be amplified by a factor of about 100  \citep[e.g.][]{2005Voelketal, 2005Vink}. This is much more than can be expected by mere compression of the circumstellar magnetic field by a factor four in a typical strong blast wave. The spectral properties of the non-thermal emission indicate that the cosmic ray diffusion must be near the Bohm-limit and that magnetic field turbulence is high (e.g.~\citet{2005Vink, 2006Stageetal}). Additionally, information on the field topology from radio-synchrotron polarization data indicates that young SNRs have a mostly radially oriented magnetic field, whereas the field in old remnants is mainly circumferential \citep{1976DickelMilne}. 

The diffusion of CRs in SNRs depends on the orientation of the mean field and the level of magnetic turbulence. In order to obtain a better understanding of the interplay between CRs and magnetic fields in SNRs, we believe it is important to also understand the structure of magnetic fields better. Although the CRs may self-generate magnetic fields \citep{1975Skilling}, for now we will treat them as test-particles and only look at the evolution of the field assuming ideal MHD. 

The contact discontinuity between the supernova ejecta and the compressed circumstellar medium, just behind the SNR blast wave, is unstable to the Rayleigh-Taylor (R-T) instability. It was suggested by \citet{1973Gull} that the radial fingers that result from this instability stretch the field lines radially, thus creating a dominant radial magnetic field component. The stretching of the field along the R-T fingers has been confirmed by simulations, for instance see \citet{1995Junetal}. While this causes the dominance of the radial field signature in part of the remnant, for a typical shock compression ratio of four, it does not explain the radial field polarization observed near the forward shock, since the R-T fingers do not extend that far outward \citep{1992Chevalieretal}.

\citet{2001BlondinEllison} have performed hydrodynamic simulations of SNRs in which they show that, when the compression ratio is higher, the R-T fingers almost reach out to the blast wave.
The presence of a cosmic ray gas with sufficient pressure would lead to an effective adiabatic index lower than that for a mono-atomic non-relativistic gas, which has $\gamma=5/3$, and thus to a higher compression ratio ($s=(\gamma+1)/(\gamma-1)$). For a gas that is dominated by relativistic particles, such as cosmic rays, the adiabatic index $\gamma=4/3$. Additionally, escaping cosmic rays drain energy from the shock region, thereby further increasing the compression ratio.

We explore the development of Rayleigh-Taylor fingers and magnetic field variations of a SNR inside a stellar wind medium with a density $\rho \propto r^{-2}$ for different initial magnetic topologies. We also investigate the influence of a softened equation of state on the properties of the resulting SNR. We compare our results with those from various studies that have been performed on this subject in a homogeneous and/or unmagnetised medium. The models for the magnetic field are the same ones we used and described by \citet{2008Schureetalb}. They will be summarised in section~\ref{sec:methods}, where we will also describe the implementation of the problem in the code we use: {\tt AMRVAC}. In line with the hydrodynamic simulations by \citet{2001BlondinEllison} we adopt a softer equation of state, and compare the differences of our MHD models for the values of $\gamma=5/3$ and $\gamma=1.1$, an extreme case that is relevant if cosmic rays are the dominant component of the plasma and escape, softening the effective equation of state to one with $\gamma < 4/3$. Our focus will be on the early evolution of the SNR, investigating if we can reproduce the observed radial field based on a choice of initial magnetic field topologies and the equation of state.

\section{Method}
\label{sec:methods}

We use the same model for the SN ejecta and circumstellar wind as that adopted in \citet{2008Schureetalb}, in which the supernova ejecta evolve in a circumstellar medium (CSM) that is shaped by a pre-supernova wind. The initial grid is filled with a RSG wind, with a mass loss rate of $\dot M = 1.54 \times 10^{-5}$~M$_\odot$~yr$^{-1}$, the wind velocity $v=4.7$~km~s$^{-1}$ (at $R\approx 0.03$~pc) and the temperature $T=1000$~K. The SN ejecta consist of a constant-density core, with an envelope for which the density decreases as $\rho \propto r^{-9}$, which is the typical density profile of the ejecta after propagation through the star in explosion models \citep[c.f.][]{1999TrueloveMcKee}. The explosion mass and energy are respectively set to be: $M_{\rm ej} = 2.5 M_\odot$ and $E_{\rm ej}=2 \times 10^{51}$~erg. The maximum velocity of the outer ejecta is chosen to be 15,000 km~s$^{-1}$.

We use the Adaptive Mesh Refinement version of the Versatile Advection Code: {\tt AMRVAC} \citep{2007HolstKeppens} to solve the MHD equations in the $r - \theta$ plane of a spherical grid, with symmetry around the polar axis. The equations that are solved are the conservative equations for mass, momentum and energy, and the induction equation:

\begin{eqnarray}
&&\partial_t \rho + \nabla \cdot ({\bf v}\rho)  =  0\\
&&\partial_t (\rho {\bf v})+\nabla \cdot ({\bf v} \rho {\bf v}) - {\bf B B})+\nabla p_{\rm tot}  =  0\\
&&\partial_t e + \nabla \cdot ({\bf v}e-{\bf BB}\cdot{\bf v}+{\bf v} p_{\rm tot})  = 0 \\
&&\partial_t {\bf B} + \nabla \cdot ({\bf v B}-{\bf Bv})  =  -(\nabla \cdot {\bf B}){\bf v}
\end{eqnarray}

The divergence of the magnetic field is controlled by adding a source term proportional to $\nabla \cdot B$ to the induction equation, while maintaining conservation of momentum and energy \citep{2003Keppensetal, 2000Janhunen}. Additionally, the initial magnetic field strength is chosen to be weak, such that it does not influence the dynamics. 

The initial grid is filled with a RSG wind, after which we introduce a magnetic field into the entire grid, and the energetic supernova ejecta into the inner 0.3 pc. 
We implement three different models for the initial magnetic field topology, which are described in \citet{2008Schureetalb} and will be summarised below: 

In Models~A and D the field is predominantly toroidal, with a small radial component, and is modelled after a rotating dipole field in the equatorial plane with a $\sin \theta$ dependence such that the toroidal field vanishes at the poles \citep{1994ChevalierLuo,1999GarciaSeguraetal}. The different components of the field decay with radius as $B_\phi \propto r^{-1}$ and $B_r \propto r^{-2}$, such that away from the star, at radial scales relevant to the SNR evolution, the toroidal field component strongly dominates.

In Model~B and E, we add a 2D turbulent magnetic field at the time when we 
introduce the ejecta. This is to mimic the presence of cosmic rays that may induce such turbulence. The azimuthal $B_\phi$ component is initially zero. The field is calculated on a 2D cartesian grid to be divergence-free, following \citet{1999GiacaloneJokipii}, and transcribed to spherical coordinates in the $r,\theta$ plane. The field is set up using the real part of:

\begin{eqnarray}
\delta {\bf B}(x,y)= \sum_{n=1}^{N_{nk}}A_0 k^{-\alpha/2}_n i( \cos \phi_n{\bf \hat y}-\sin\phi_n{\bf \hat x}) \\\nonumber \times \quad e^{i k_n(x\cos\phi_n+y\sin\phi_n)+i \beta_n},
\end{eqnarray}
with $N_{nk}=256$ different modes with wavenumbers $k$. The phase $\beta$ and polarization $\phi$ is randomly chosen between $0$ and $2\pi$ for each wavenumber. The wavenumbers are logarithmically spaced and span the range between a wavelength corresponding to two gridcells, and a wavelength that covers the entire grid. We approximate a Kolmogorov spectrum by adopting $\alpha=5/3$.

In Model~C and F, the magnetic field is taken to be unidirectional and parallel, representing a situation like the dominant ordered magnetic field as observed in e.g. spiral galaxies. We choose the direction of the field to be alighned with the symmetry axis. This magnetic field topology may in reality be more appropriate to model type Ia SNe instead of the ones that explode in a stellar-wind environment. In spherical coordinates, this uniform magnetic field is represented by $B_r = B \cos \theta$, $B_\theta = -B \sin \theta$.

Models A-C are set up with an adiabatic index of $\gamma=5/3$, the value for a monatomic gas. The grid spans $1.8 \times 10^{19}$~cm radially, and an effective resolution of $2.75 \times 10^{15}$~cm by $0.125^\circ$ is reached in regions where strong density and/or velocity gradients are present. When cosmic ray pressure dominates the fluid, the value of the adiabatic index drops, approaching the value $\gamma=4/3$ for a relativistically hot gas. Because of escape of cosmic rays the equation of state can soften even further. In this paper we evaluate the extreme case where $\gamma=1.1$ in models D-F. In those simulations, the grid spans $3.0 \times 10^{19}$~cm radially and the effective resolution reaches $6.25 \times 10^{15}$~cm by $0.125^\circ$.

\section{Results}
\label{sec:results}

\begin{figure}[!htbp]
   \centering
     \includegraphics[width=\columnwidth]{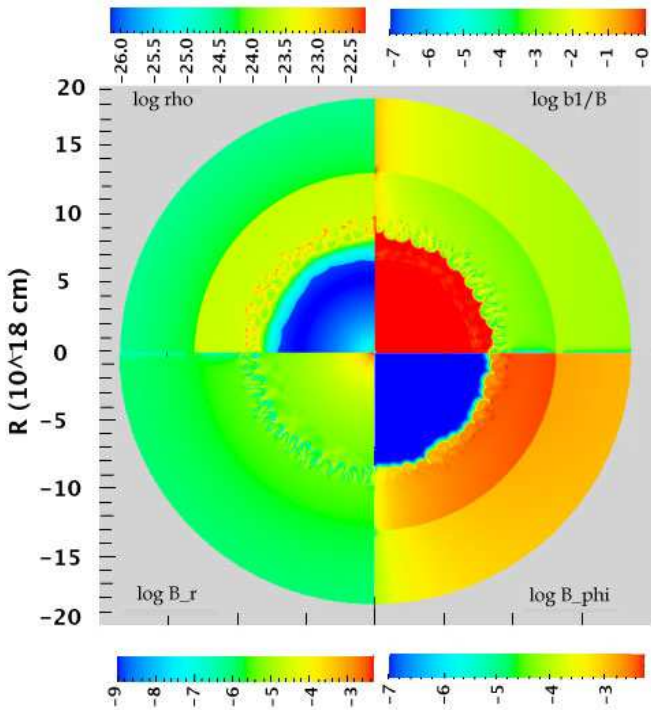}
      \caption{Simulation of SNR into CSM with a mostly {\em toroidal} initial field (Model~A) and an adiabatic index of $\gamma=5/3$, shown at a time of 634 yr after explosion. The upper left quadrant shows the logarithm of the density, which shows clearly the occurence of the Rayleigh-Taylor instability at the contact discontinuity. The density jumps at a radius of $1.3 \times 10^{19}$~cm and $\sim 7.0 \times 10^{18}$~cm correspond to the locations of the forward and the reverse shock respectively. The lower left panel shows the logarithm of the absolute value of the radial magnetic field. The lower right part shows the logarithm of $|B_\phi|$, and the upper right quadrant shows the logarithm of the ratio of the radial field component relative to the total field ($|B_r|/|{\bf B}|$).}
         \label{fig:modelA}
   \end{figure}

\begin{figure}[!htbp]
   \centering
     \includegraphics[width=\columnwidth]{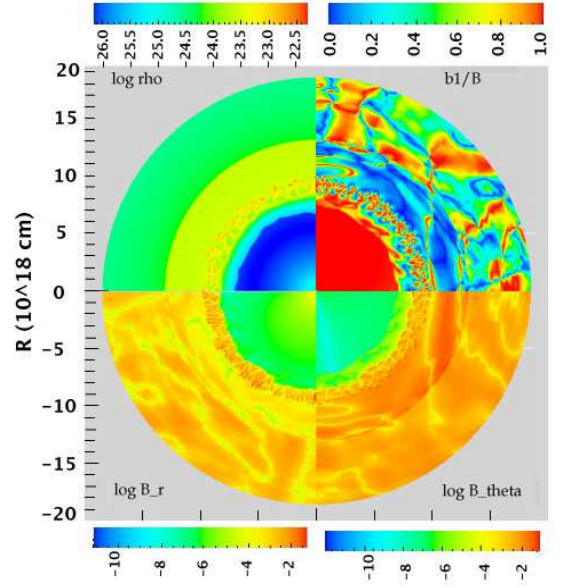}
      \caption{Simulation of the SNR into CSM with a {\em turbulent} initial field (Model~B). Similar to Figure~\ref{fig:modelA}, the density and magnetic field is plotted. Note that in this case the lower right part shows the logarithm of $|B_\theta|$, and the upper right quadrant shows the ratio of the radial field component relative to the total field ($|B_r|/|{\bf B}|$).}
         \label{fig:modelB}
   \end{figure}

\begin{figure}[!htbp]
   \centering
     \includegraphics[width=\columnwidth]{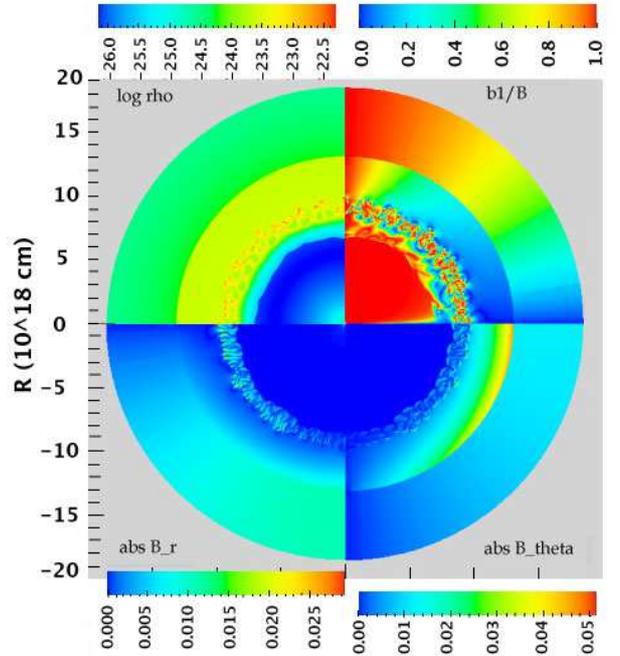}
      \caption{Simulation of SNR into CSM with an initial magnetic field {\em parallel} to the symmetry axis, i.e. vertically aligned (Model~C). The upper quadrants show again the logarithm of the density and the ratio of radial field over the total field. The lower quadrants show the absolute values of $B_r$ and $B_\theta$. }
         \label{fig:modelC}
   \end{figure}

The results from simulations of Models A, B and C, are plotted in Figures~\ref{fig:modelA} to \ref{fig:modelC}. The density, radial and toroidal magnetic field, and additionally the ratio of the radial to the total field are plotted for the three assumed initial magnetic-field topologies. The results shown here are for remnants that have evolved for a period of 634~year.

%
%
%

The supernova ejecta sweeps up the circumstellar matter into a dense shell. Four distinct regions can be identified: the unshocked circumstellar medium (CSM) ahead of the blast wave, the shocked CSM, the shocked ejecta, and the freely expanding ejecta. The shocked CSM is separated from the unshocked CSM by a strong shock (in Figures~\ref{fig:modelA}-\ref{fig:modelC} at a radius of $R_{\rm f} \approx 1.3 \times 10^{19}$~cm), characterised by a pressure jump, which increases the density by a factor $(\gamma+1)/(\gamma-1)$ and decreases the velocity in the shock frame by the same amount. The shocked CSM is separated from the third region, the shocked ejecta, by the contact discontinuity, characterised by a jump in the density, but at constant pressure. This is the location where the R-T instability develops. The reverse shock, at a radius of $R_{\rm r} \approx 6.5 \times 10^{18}$~cm, marks the boundary with the unshocked ejecta. 

The deceleration of the shocked ejecta by the less-dense shocked CSM is Rayleigh-Taylor unstable. Since the magnetic field is very weak, it does not influence the dynamics of the blast wave, and we do not see a difference between the three models in the development of the R-T instabilities and the propagation of the blast wave. The  compression ratio at the forward shock is too low to allow for the Rayleigh-Taylor fingers to extend out to the forward shock, in agreement with findings by \citet{1992Chevalieretal}. \citet{1995ChevalierBlondin} found similar limitations to the extent of the R-T fingers when radiative cooling was taken into account.

The magnetic field is carried along by the plasma. Since the velocity field initially only has a radial component, the induction equation ($\partial_t {\bf B} = \nabla \times ({\bf v} \times {\bf B})$) implies that the toroidal field is swept up, while the radial field is just attenuated proportional to the square of the radius. In our models, the ejecta are initially set up using the same configuration for the magnetic field as is used for the CSM. As a result, the radial-field component is the dominant component in the ejecta. 
Starting at the R-T unstable region at the contact discontinuity, a sizable toroidal field is present. The R-T instability induces a $v_\theta$ component, which in turn modifies the radial component of the magnetic field. Along the R-T fingers, for Models B and C, the field is mostly radial  with toroidal field components at the tips and bases of the fingers, in agreement with results from e.g. \citet{1995Junetal}. In Model~A however, we do not see a significant radial component in the R-T unstable region, which may have to do with the small scale of the initial radial field. This case requires further investigation, allowing for a gradient in the third dimension.

Models D, E, and F correspond to models A through C, but with an adiabatic index equal to $\gamma=1.1$. The compression  ratio is analytically given by $s=(\gamma+1)/(\gamma-1)$ and therefore higher for a lower value of $\gamma$. As a consequence, the forward shock propagates more slowly into the CSM. At a time of $634$~yr, the blast wave has only reached a distance of about $1.1 \times 10^{19}$~cm. For clarity purposes we show the SNR at a later stage, at an age of approximately 1200 years. As a result of the higher compression ratio the forward and reverse shock are closer together and the R-T fingers almost reach the forward shock. The magnetic field in models E and F again is obviously stretched around the R-T fingers, resulting in a dominantly radial field over the entire remnant. The magnetic field in model D, where the field was initially mainly toroidal, does not show a preferred radial component, even around the R-T fingers. 
   \begin{figure}[!htbp]
   \centering
     \includegraphics[width=\columnwidth]{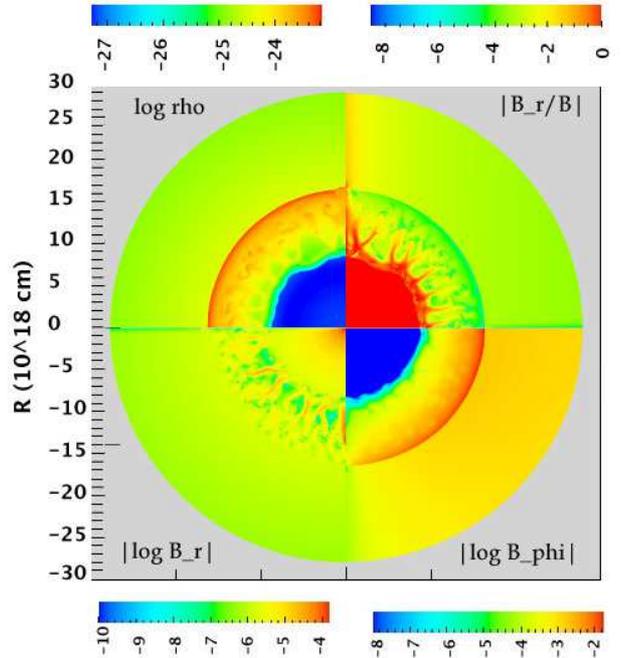}
      \caption{The SNR evolution is shown at a time of ~1200 yrs after explosion. The initial magnetic field was mainly {\em toroidal}, and the adiabatic index $\gamma=1.1$ (Model~D). The soft equation of state causes the relative proximity of forward to reverse shock, and the Rayleigh-Taylor fingers stretch almost out to the blast wave, seen best in the upper left quadrant, which shows the logarithm of the density. The lower left panel shows the logarithm of the absolute value of the radial magnetic field. The lower right part shows the logarithm of $|B_\phi|$, and the upper right quadrant shows the logarithm of the ratio of the radial field component relative to the total field ($|B_r|/|{\bf B}|$).}
         \label{fig:modelD}
   \end{figure}
%
   \begin{figure}[!htbp]
   \centering
     \includegraphics[width=\columnwidth]{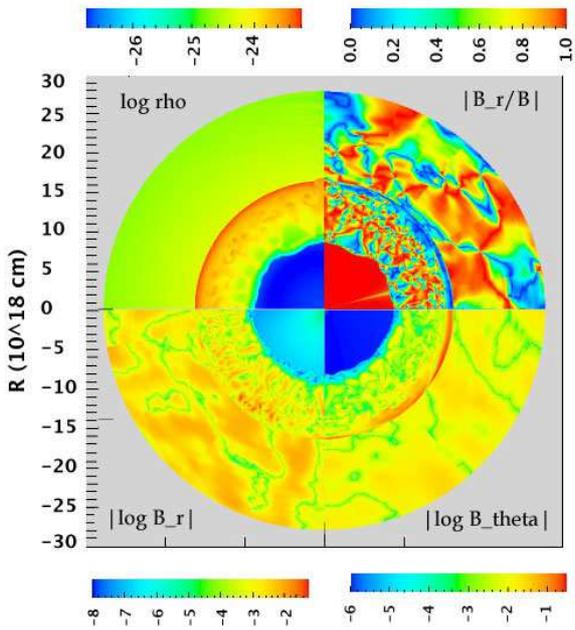}
      \caption{Simulation of the SNR into CSM with a {\em turbulent} initial field and adiabatic index of $\gamma=1.1$ (Model~E). The four panels represent the same as those in Figure~\ref{fig:modelD}, except for the lower right one, which represents the logarithm of $|B_\theta|$. The magnetic field is stretched radially around the R-T fingers, and reaches almost out to the forward shock.}
         \label{fig:modelE}
   \end{figure}
%
   \begin{figure}[!htbp]
   \centering
     \includegraphics[width=\columnwidth]{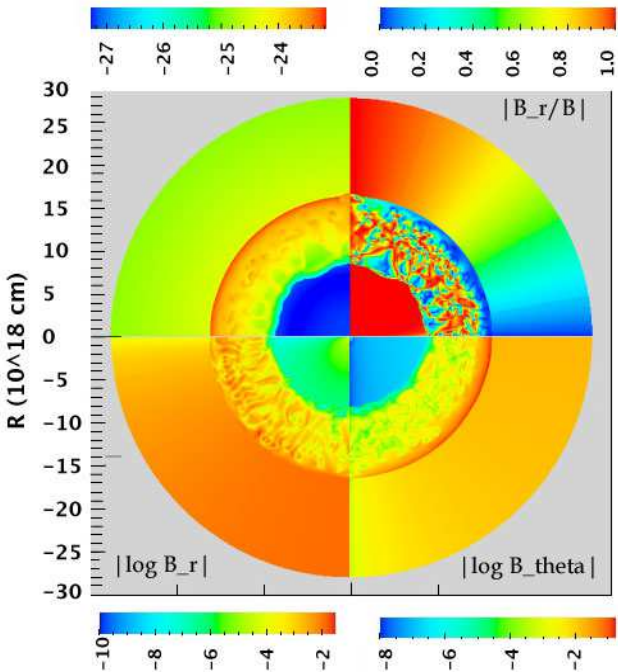}
      \caption{Simulation of SNR into CSM with an initial magnetic field parallel to the symmetry axis, i.e. vertically aligned (Model~F). The four panels represent the same variables as in Figure~\ref{fig:modelE}. Similar to the case in model~E, the magnetic field is radially dominant throughout the entire remnant, except for a small zone at the shock front itself.}
         \label{fig:modelF}
   \end{figure}
%

\section{Particle acceleration}

Observational evidence suggests that cosmic rays are efficiently accelerated at the forward shock, and possibly also at the reverse shock \citep{2008HelderVink}, up to energies of $10^{15}$~eV. The mechanism that is believed to be responsible for this is diffusive shock acceleration (DSA), also know as first-order Fermi acceleration. In this mechanism, the particle momentum increases every time the particle goes from upstream to downstream and back due to the randomization of the momentum vector by scattering in the rest frames of the up- and downstream flow. The simplest case of advection and diffusion of cosmic rays in 1D plane parallel geometry can be described by the following set of two stochastic difference equations \citep{1992AchterbergKruells, 2004vanderSwaluwAchterberg}: 

\begin{eqnarray}
du=- \frac{dV}{3 dx} dt\\
dx= V dt + \sqrt{2 \kappa dt} \xi.
\end{eqnarray}
The first equation represents the boost in energy (with $u=\ln(p/mc)$), and the second one describes the random walk of the particle, with an advective term and a diffusive term. The diffusive term contains the variable $\xi$, with $\langle \xi_i \rangle=0$ and  $\langle \xi_i \xi_j \rangle=\delta_{ij}$, to represent the stochastic behaviour of the diffusion, of which the sum of many instances returns the distribution function of the relativistic particles. The diffusion of the particle is caused by inelastic scattering by a turbulent magnetic field, described by the diffusion parameter $\kappa$. It depends on the energy of the particle, and the magnetic field strength, where diffusion parallel to the field may differ from the diffusion perpendicular to the field. In future work we will include this dependence, but for now we fix the diffusion parameter to a constant, in order to compare the spectrum to the analytical solution. The diffusive lengthscale $x_{\rm diff}=\sqrt{2 \kappa dt}$ needs to be larger than the shock thickness $x_{\rm shock}$ in order for the particles to feel the effects of DSA. The shock is spread over approximately 5 grid cells, which sets a lower limit to the diffusion parameter, depending on the resolution. Additionally the advection length scale  $x_{\rm adv}=V dt$ should be smaller than the shock thickness, for the particles to 'see' the shock.
The condition that needs to be satisfied is

\begin{eqnarray}
\label{eq:deltax}
\Delta x_{adv} < \Delta x_{shock} \ll \Delta x_{diff}.
\end{eqnarray}

If these conditions are met, the resulting energy spectrum should approach the analytical power-law solution of a strong shock, where slope $q$ of the power-law is determined by the shock compression ratio $s=(\gamma+1)/(\gamma-1)$. 
One has \citep{1992AchterbergKruells}:

\begin{eqnarray}
F(u) \propto p\frac{dN}{dp} \propto p^{-q+1},
\end{eqnarray}
with
\begin{eqnarray}
q=(s+2)/(s-1).
\end{eqnarray}

\begin{figure}[!htbp]
   \centering
     \includegraphics[width=\columnwidth]{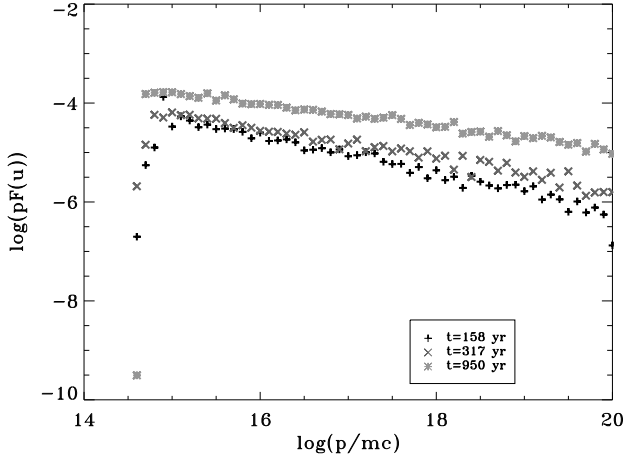}
      \caption{Cosmic ray spectrum for $\gamma=5/3$, at different points in time of the SNR evolution.}
         \label{fig:spectrum}
   \end{figure}

In figure~\ref{fig:spectrum} we show the spectrum for the case where $\gamma=5/3$, implicating $s=4$ and $q=2$. The coordinate system is set to slab geometry, such that no adiabatic losses occur.  
The mass and energy of the supernova explosion are respectively $2.5$~M$_\odot$ and $2.0 \times 10^{51}$~erg and the circumstellar medium is taken to be homogeneous with a density of $1.64 \times 10^{-21}$~g~cm$^{-3}$. Particles are introduced at the forward shock position, with a fixed initial energy, and the total number of particles increases linearly with time. 

The resulting spectrum for different times in the SNR evolution is plotted in Figure~\ref{fig:spectrum}. For the higher energy particles it takes longer to reach the equilibrium value of the spectrum that can be compared with the analytical solution, and we can see the evolution of the spectrum towards higher energies for later times. In the plane parallel geometry that we assume, there is no theoretical limit to the maximum energy the particles can attain. In spherical geometry, in addition to the time during which particles have been accelerated, the maximum energy is limited by adiabatic losses and the size of the remnant \citep[see e.g.][]{1983Drury,2000Achterberg, 2001BellLucek}. DSA in spherical geometry will be treated and discussed in a future work.

%

\section{Conclusions and Discussion}

For an ideal mono-atomic gas that is not significantly influenced by the presence of cosmic rays, our models show that the radial magnetic field does not dominate in the outer region of the remnant, contrary to observations. For a softer equation of state, such as may be induced by cosmic rays if particle acceleration is effective in the SNR, the field can actually become radial for the entire remnant, apart from a thin layer at the forward shock. This is seen in models~E and F with the parallel and turbulent initial magnetic field topologies.
Since cosmic rays are believed to randomise the magnetic field near the shock in supernova remnants through streaming instabilities, the scenario where a magnetic field is initially turbulent and the equation of state is softer than $\gamma=5/3$ is a realistic possiblity \citep{2001BlondinEllison}. Whether or not the equation of state can become even softer than that of a relativistic gas, where $\gamma=4/3$, depends on the slope of the spectrum. For a hard spectrum ($q<2$), high energy particles that escape from the system can carry away a substantial fraction of the energy, thus lowering the effective adiabatic index. Observational evidence so far, favours a softer spectrum ($q>2$), in which case the compression ratio would be limited to up to about $7$ for $\gamma=4/3$. Models to explain the observed H.E.S.S. spectrum in $\gamma$-rays of the SNR RX J1713.7-3946 by \citet[e.g.][]{2006BerezhkoVoelk, 2008BerezhkoVoelk} that include nonlinear feedback between the CRs and the hydrodynamics, require a compression ratio of $s=6.3$ in order to explain the observed spectrum. In such a case, the R-T fingers would not reach the forward shock.
Although the toroidal field as used in models~A and D is the favoured one for the magnetic field in stellar winds, this does not reproduce the radial field orientation from observations. It is possible however, that even in such a geometry, the cosmic ray streaming creates enough turbulence to effectively create a scenario like we use in model E. Even though further upstream the field may be toroidal, in the direct vicinity of the shock front and downstream the field may be turbulent because of the cosmic rays, and consequently become mainly radial around the R-T fingers. 
The development of the radial field may be suppressed by the limitations of the 2.5D model, for which, because of the lack of a gradient in the $\phi$-direction, the induction equation lacks a number of terms. This effect is present for all three models described here. Full 3D simulations are needed to see if the results presented here are typical, including the possible difference in the development of R-T instabilities in 3D compared to the 2.5D results presented here.

An alternative scenario to explain the radial magnetic field orientation near the blast wave was proposed by \citet{2008ZirakashviliPtuskin}. They showed that the non-resonant streaming instability can amplify and radially stretch the magnetic field downstream of the shock front, when small density perturbations naturally create enough turbulence to trigger this effect in SNRs. The radial field can thus be a factor of 1.4 larger than the component parallel to the shock front.

Quantitatively, it is very complicated to make realistic predictions about values for the adiabatic index and level of turbulence. Since magnetic fields are additionally amplified at the shock \citep[e.g.][]{2001BellLucek}, the magnetic pressure may become an additional non-negligible component, again changing the compression ratio at the shock as well as the pressure profile. This was not taken into account in our models.
In the test particle approach, the distribution of accelerated particles should follow a power law with a slope that is dependent on the adiabatic index. Preliminary results of our implementation of test particles and dynamical evolution with the MHD itself are promising. Nonlinear feedback of the particle acceleration to the magnetic field and the equation of state are expected to be important. In future studies we hope to further evaluate the interaction and the resulting particle spectrum and observational features that it will produce.

\bigskip\noindent
{\small
This study has been financially supported by J.V.'s Vidi grant from the Netherlands Organisation for Scientific Research (NWO). This work was sponsored by the Stichting Nationale Computerfaciliteiten (National Computing Facilities Foundation, NCF) for the use of supercomputer facilities, with financial support from the Nederlandse Organisatie voor Wetenschappelijk Onderzoek (Netherlands Organization for Scientific Research, NWO). K.M.S. acknowledges the hospitality of the astronomy department at the University of Florida, where part of this work was performed.     
}

\bibliography{adssamplecospar}

\begin{thebibliography}{27}
\expandafter\ifx\csname natexlab\endcsname\relax\def\natexlab#1{#1}\fi
\expandafter\ifx\csname url\endcsname\relax
  \def\url#1{\texttt{#1}}\fi
\expandafter\ifx\csname urlprefix\endcsname\relax\def\urlprefix{URL }\fi

\bibitem[{{Achterberg}(2000)}]{2000Achterberg}
{Achterberg}, A., May 2000. {Particle Acceleration at Astrophysical Shocks}.
  In: {Martens}, P.~C.~H., {Tsuruta}, S., {Weber}, M.~A. (Eds.), Highly
  Energetic Physical Processes and Mechanisms for Emission from Astrophysical
  Plasmas. Vol. 195 of IAU Symposium. pp. 291--+.

\bibitem[{{Achterberg} and {Krulls}(1992)}]{1992AchterbergKruells}
{Achterberg}, A., {Krulls}, W.~M., Nov. 1992. {A fast simulation method for
  particle acceleration}. A\&A 265, L13--L16.

\bibitem[{{Bell} and {Lucek}(2001)}]{2001BellLucek}
{Bell}, A.~R., {Lucek}, S.~G., Mar. 2001. {Cosmic ray acceleration to very high
  energy through the non-linear amplification by cosmic rays of the seed
  magnetic field}. MNRAS 321, 433--438.

\bibitem[{{Berezhko} and {V{\"o}lk}(2006)}]{2006BerezhkoVoelk}
{Berezhko}, E.~G., {V{\"o}lk}, H.~J., Jun. 2006. {Theory of cosmic ray
  production in the supernova remnant RX J1713.7-3946}. A\&A 451, 981--990.

\bibitem[{{Berezhko} and {V{\"o}lk}(2008)}]{2008BerezhkoVoelk}
{Berezhko}, E.~G., {V{\"o}lk}, H.~J., 2008. {Comparing a model of cosmic ray
  production in the supernova remnant RX J1713.7-3946 with observations}. In:
  International Cosmic Ray Conference. Vol.~2 of International Cosmic Ray
  Conference. pp. 259--262.

\bibitem[{{Blondin} and {Ellison}(2001)}]{2001BlondinEllison}
{Blondin}, J.~M., {Ellison}, D.~C., Oct. 2001. {Rayleigh-Taylor Instabilities
  in Young Supernova Remnants Undergoing Efficient Particle Acceleration}. ApJ
  560, 244--253.

\bibitem[{{Chevalier} and {Blondin}(1995)}]{1995ChevalierBlondin}
{Chevalier}, R., {Blondin}, J.~M., May 1995. {Hydrodynamic instabilities in
  supernova remnants: Early radiative cooling}. ApJ 444, 312--317.

\bibitem[{{Chevalier} et~al.(1992){Chevalier}, {Blondin}, and
  {Emmering}}]{1992Chevalieretal}
{Chevalier}, R.~A., {Blondin}, J.~M., {Emmering}, R.~T., Jun. 1992.
  {Hydrodynamic instabilities in supernova remnants - Self-similar driven
  waves}. ApJ 392, 118--130.

\bibitem[{{Chevalier} and {Luo}(1994)}]{1994ChevalierLuo}
{Chevalier}, R.~A., {Luo}, D., Jan. 1994. {Magnetic shaping of planetary
  nebulae and other stellar wind bubbles}. ApJ 421, 225--235.

\bibitem[{{Dickel} and {Milne}(1976)}]{1976DickelMilne}
{Dickel}, J.~R., {Milne}, D.~K., Oct. 1976. {Magnetic fields in supernova
  remnants}. Australian Journal of Physics 29, 435--460.

\bibitem[{{Drury}(1983)}]{1983Drury}
{Drury}, L.~O., Aug. 1983. {An introduction to the theory of diffusive shock
  acceleration of energetic particles in tenuous plasmas}. Reports on Progress
  in Physics 46, 973--1027.

\bibitem[{{Garc{\'{\i}}a-Segura} et~al.(1999){Garc{\'{\i}}a-Segura}, {Langer},
  {R{\'o}{\.z}yczka}, and {Franco}}]{1999GarciaSeguraetal}
{Garc{\'{\i}}a-Segura}, G., {Langer}, N., {R{\'o}{\.z}yczka}, M., {Franco}, J.,
  Jun. 1999. {Shaping Bipolar and Elliptical Planetary Nebulae: Effects of
  Stellar Rotation, Photoionization Heating, and Magnetic Fields}. ApJ 517,
  767--781.

\bibitem[{{Giacalone} and {Jokipii}(1999)}]{1999GiacaloneJokipii}
{Giacalone}, J., {Jokipii}, J.~R., Jul. 1999. {The Transport of Cosmic Rays
  across a Turbulent Magnetic Field}. ApJ 520, 204--214.

\bibitem[{{Gull}(1973)}]{1973Gull}
{Gull}, S.~F., 1973. {A numerical model of the structure and evolution of young
  supernovaremnants}. MNRAS 161, 47--+.

\bibitem[{{Helder} and {Vink}(2008)}]{2008HelderVink}
{Helder}, E.~A., {Vink}, J., Oct. 2008. {Characterizing the Nonthermal Emission
  of Cassiopeia A}. ApJ 686, 1094--1102.

\bibitem[{{Janhunen}(2000)}]{2000Janhunen}
{Janhunen}, P., May 2000. {A Positive Conservative Method for
  Magnetohydrodynamics Based on HLL and Roe Methods}. J. Comput. Phys. 160~(2),
  649--661.

\bibitem[{{Jun} et~al.(1995){Jun}, {Norman}, and {Stone}}]{1995Junetal}
{Jun}, B.-I., {Norman}, M.~L., {Stone}, J.~M., Nov. 1995. {A Numerical Study of
  Rayleigh-Taylor Instability in Magnetic Fluids}. ApJ 453, 332--+.

\bibitem[{{Keppens} et~al.(2003){Keppens}, {Nool}, {T\'oth}, and
  {Goedbloed}}]{2003Keppensetal}
{Keppens}, R., {Nool}, M., {T\'oth}, G., {Goedbloed}, J.~P., Jul. 2003.
  {Adaptive Mesh Refinement for conservative systems: multi-dimensional
  efficiency evaluation}. CoPhC 153, 317--339.

\bibitem[{Schure et~al.(2008)Schure, Vink, Achterberg, and
  Keppens}]{2008Schureetalb}
Schure, K.~M., Vink, J., Achterberg, A., Keppens, R., Oct. 2008. {Evolution of
  Magnetic Fields in Supernova Remnants}. ArXiv e-prints.

\bibitem[{{Skilling}(1975)}]{1975Skilling}
{Skilling}, J., Nov. 1975. {Cosmic ray streaming. II - Effect of particles on
  Alfven waves}. MNRAS 173, 245--254.

\bibitem[{{Stage} et~al.(2006){Stage}, {Allen}, {Houck}, and
  {Davis}}]{2006Stageetal}
{Stage}, M.~D., {Allen}, G.~E., {Houck}, J.~C., {Davis}, J.~E., Sep. 2006.
  {Cosmic-ray diffusion near the Bohm limit in the Cassiopeia A supernova
  remnant}. Nature Physics 2, 614--619.

\bibitem[{{Truelove} and {McKee}(1999)}]{1999TrueloveMcKee}
{Truelove}, J.~K., {McKee}, C.~F., Feb. 1999. {Evolution of Nonradiative
  Supernova Remnants}. ApJS 120, 299--326.

\bibitem[{{van der Holst} and {Keppens}(2007)}]{2007HolstKeppens}
{van der Holst}, B., {Keppens}, R., Sep. 2007. {Hybrid block-AMR in cartesian
  and curvilinear coordinates: MHD applications}. J. Comp. Phys. 226, 925--946.

\bibitem[{{van der Swaluw} and {Achterberg}(2004)}]{2004vanderSwaluwAchterberg}
{van der Swaluw}, E., {Achterberg}, A., Jul. 2004. {Non-thermal X-ray emission
  from young supernova remnants}. A\&A 421, 1021--1030.

\bibitem[{{Vink}(2005)}]{2005Vink}
{Vink}, J., Feb. 2005. {Non-thermal X-ray Emission from Supernova Remnants}.
  AIP Conf. Proc. 745, 160--171.

\bibitem[{{V{\"o}lk} et~al.(2005){V{\"o}lk}, {Berezhko}, and
  {Ksenofontov}}]{2005Voelketal}
{V{\"o}lk}, H.~J., {Berezhko}, E.~G., {Ksenofontov}, L.~T., Apr. 2005.
  {Magnetic field amplification in Tycho and other shell-type supernova
  remnants}. A\&A 433, 229--240.

\bibitem[{{Zirakashvili} and {Ptuskin}(2008)}]{2008ZirakashviliPtuskin}
{Zirakashvili}, V.~N., {Ptuskin}, V.~S., May 2008. {Diffusive Shock
  Acceleration with Magnetic Amplification by Nonresonant Streaming Instability
  in Supernova Remnants}. ApJ 678, 939--949.

\end{thebibliography}

\end{document}